\documentclass[12pt,preprint]{aastex}
\usepackage{color}
\usepackage{graphics,graphicx}
\usepackage{rotating}

\newcommand{\lapprox} {\, \lower3pt\hbox{$\sim$}\llap{\raise2pt\hbox{$<$}}\,}
\newcommand{\gapprox} {\, \lower3pt\hbox{$\sim$}\llap{\raise2pt\hbox{$>$}}\,}

\begin{document}

\title{PROPERTIES OF THE ACCELERATION REGIONS IN SEVERAL LOOP-STRUCTURED SOLAR FLARES}
\author{Jingnan~Guo\altaffilmark{1},
        A.~Gordon~Emslie\altaffilmark{2},
        Anna~Maria~Massone\altaffilmark{3},
        AND Michele~Piana\altaffilmark{1,3}}

\altaffiltext{1}{Dipartimento di Matematica, Universit\`a di Genova, via Dodecaneso 35, 16146 Genova, Italy; guo@dima.unige.it, piana@dima.unige.it}

\altaffiltext{2}{Department of Physics and Astronomy, Western Kentucky University, Bowling Green, KY 42101; emslieg@wku.edu}

\altaffiltext{3}{CNR - SPIN, via Dodecaneso 33, I-16146 Genova, Italy; annamaria.massone@cnr.it}

\begin{abstract}

Using {\em RHESSI} hard X-ray imaging spectroscopy observations, we analyze electron flux maps for a number of extended coronal loop flares. For each event, we fit a collisional model with an extended acceleration region to the observed variation of loop length with electron energy $E$, resulting in estimates of the plasma density in, and longitudinal extent of, the acceleration region.  These quantities in turn allow inference of the number of particles within the acceleration region and hence the filling factor $f$ -- the ratio of the emitting volume to the volume that encompasses the emitting region(s).  We obtain values of $f$ that lie mostly between $0.1$ and $1.0$; the (geometric) mean value is $f = 0.20 \, \times\!/\!\div \, 3.9$, somewhat less than, but nevertheless consistent with, unity.  Further, coupling information on the number of particles in the acceleration region with information on the total rate of acceleration of particles above a certain reference energy (obtained from spatially-integrated hard X-ray data) also allows inference of the specific acceleration rate (electron~s$^{-1}$ per ambient electron above the chosen reference energy). We obtain a (geometric) mean value of the specific acceleration rate $\eta(20$~keV) $ = (6.0 \, \times\!/\!\!\div \, 3.4) \times 10^{-3}$~electrons~s$^{-1}$~per ambient electron; this value has implications both for the global electrodynamics associated with replenishment of the acceleration region and for the nature of the particle acceleration process.

\end{abstract}


\section{Introduction}\label{intro}

An important diagnostic of high-energy electrons accelerated in solar flares is the hard X-ray bremsstrahlung that they produce as they propagate through the ambient solar atmosphere.  The Ramaty High Energy Solar Spectroscopic Imager ({\em RHESSI\ }) has revealed a new class of flares in which the bulk of the hard X-ray emission is produced predominantly not in dense chromospheric footpoints, but rather in the coronal loop \citep{vebr04, suetal04,krucker08}.  For such sources, the corona is not only the site of particle acceleration, but also dense enough to act as a thick target, stopping the accelerated electrons before they can penetrate to the chromosphere.

For suprathermal electrons with energy substantially greater than the thermal energy of the ambient electrons with which they interact, it is appropriate to use a collisional cold-target energy loss rate \citep[e.g.,][]{1978ApJ...224..241E}, for which the penetration depth of electrons increases with energy. \citet{xuetal08} analyzed a set of extended coronal flare loops located near the solar limb, and were indeed able to account for the observed behavior of loop extent with photon energy $\epsilon$ in terms of a cold-target collisional model with an extended acceleration region. \citet{guoetal2012} have extended this analysis technique to a study of the variation of loop size with {\it electron} energy $E$, in which the visibilities used to construct the electron flux images are obtained by regularized spectral inversion of the visibility data in the count domain \citep{pianaetal07}.

Here we apply this new analysis technique to several simple coronal loop events observed by {\em RHESSI}.  In Section~\ref{events}, we present basic data for the 22 events used in the study.  In Section~\ref{analysis} we fit the variation of loop size with electron energy $E$ to the parametric model of \citet{guoetal2012} in order to determine the acceleration region length $L_0$ and density $n$ for each event.  In Section~\ref{results} these values are used to determine estimates of two important properties of the acceleration region -- the filling factor $f$ (the ratio of the volume that is actively involved in electron acceleration to the overall volume that encompasses the acceleration region[s]) and the specific acceleration rate $\eta(E_0)$ (the rate of acceleration of electrons to energies $\ge E_0$ per ambient electron), and we compare the values of these quantities to the predictions of various acceleration models.

\section{Events Studied}\label{events}

\begin{deluxetable} {cccccccc}
    \tablewidth{0pt}
    \tabletypesize{\scriptsize}
    \tablecaption{Event List and Spectral Fit Parameters\label{table:spec}}
    \tablehead{
    \colhead{Event No.} & \colhead{Date} & \colhead{Time (UT)} & \colhead{EM ($10^{49}$~cm$^{-3}$)} & \colhead{T (keV) } & \colhead{$\delta$} & \colhead{$E_t$ (keV)} & \colhead{$d{\cal N}/dt$ ($10^{35}$~s$^{-1}$) } }
    \startdata
	1 & 2002-04-12 & 17:42:00-17:44:32 & $0.30$ & $1.53$ & $8.24$ & $15.5$ & $2.71$ \\
	 2 & 		   & 17:45:32-17:48:00 & $0.46$ & $1.54$ & $8.01$ & $15.5$ & $4.69$ \\
	  \hline
 	3 & 2002-04-15 & 00:00:00-00:05:00 & $0.22$ & $1.75$ & $7.48$ & $15.5$ & $4.70$ \\
	 4 & 		   & 00:05:00-00:10:00 & $0.76$ & $1.61$ & $7.93$ & $15.5$ & $9.32$ \\
	 5 & 		   & 00:10:00-00:15:00 & $1.02$ & $1.60$ & $8.37$ &$15.5$ & $11.41$ \\
	  \hline
 	6 & 2002-04-17 & 16:54:00-16:56:00 & $0.06$ & $1.51$ & $5.70$ & $15.5$ & $0.39$ \\
	 7 & 		   & 16:56:00-16:58:00 & $0.22$ & $1.43$ & $8.78$ & $14.8$ & $2.43$ \\
	  \hline
 	8 & 2003-06-17 & 22:46:00-22:48:00 & $1.92$ & $1.71$ & $9.95$ &$16.5$ & $17.27$ \\
	 9 & 		   & 22:48:00-22:50:00 & $2.59$ & $1.67$ & $10.36$ &$16.5$ & $17.91$ \\
	  \hline
 	10 & 2003-07-10 & 14:14:00-14:16:00 & $1.26$ & $1.45$ & $10.05$ & $15.5$ & $7.43$ \\
	 11 & 		   & 14:16:00-14:18:00 & $1.31$ & $1.34$ & $10.38$ & $14.8$ & $8.53$ \\
	  \hline
 	12 & 2004-05-21 & 23:47:00-23:50:00 & $0.35$ & $1.85$ & $7.07$ & $18.5$ & $3.28$ \\
	 13 & 		   & 23:50:00-23:53:00 & $0.62$ & $1.75$ & $7.51$ & $18.5$ & $2.32$ \\
	  \hline
 	14 & 2004-08-31 & 05:31:00-05:33:00 & $0.06$ & $1.61$ & $10.56$ & $15.5$ & $0.40$ \\
	 15 & 		   & 05:33:00-05:35:00 & $0.21$ & $1.57$ & $12.19$ & $18.5$ & $0.29$ \\
	 16 & 		   & 05:35:00-05:37:00 & $0.29$ & $1.48$ & $7.45$  & $18.5$ & $0.16$ \\
	  \hline
 	17 & 2005-06-01 & 02:40:20-02:42:00 & $0.14$ & $1.81$ & $6.53$ & $17.5$ & $1.44$ \\
	18  & 		   & 02:42:00-02:44:00 & $0.37$ & $1.70$ & $7.86$ & $17.5$ & $2.67$ \\
	  \hline
 	19 & 2011-02-13 & 17:33:00-17:34:00 & $0.54$ & $1.39$ & $5.86$ & $10.5$ & $35.02$ \\
	 20 & 		   & 17:34:00-17:35:00 & $0.52$ & $1.68$ & $6.55$ & $14.5$ & $19.43$ \\
	  \hline
 	21 & 2011-08-03 & 04:31:12-04:33:00 & $0.36$ & $1.61$ & $9.23$ & $15.5$ & $3.96$ \\
 	\hline
 	22 & 2011-09-25 & 03:30:36-03:32:00 & $0.13$ & $1.44$ & $8.33$ & $14.5$ & $1.19$ \\
 	\hline
\enddata
\end{deluxetable}

\begin{center}
\begin{figure}[pht]
\begin{tabular}{cc}
{  \includegraphics[width=0.45\textwidth]{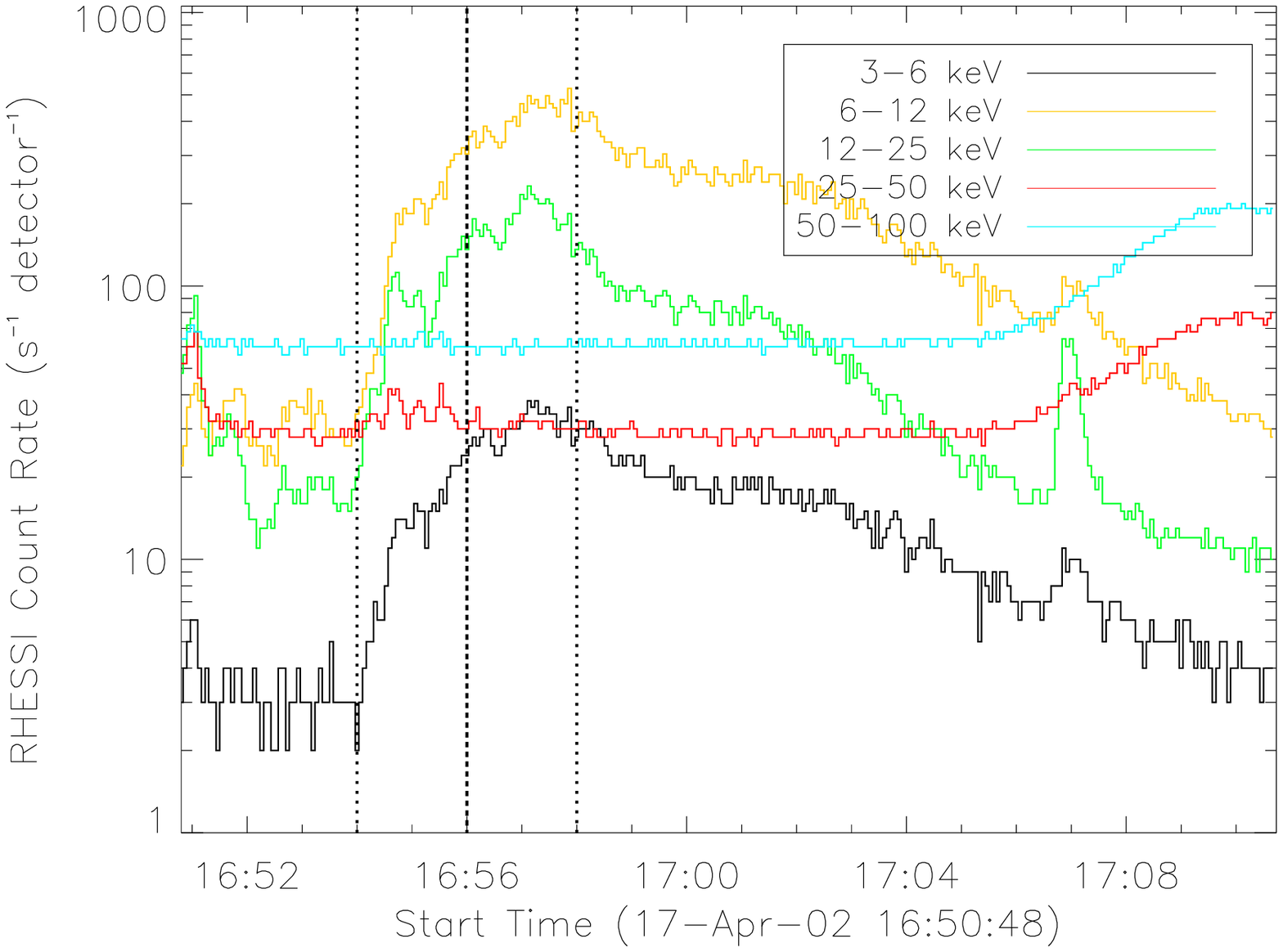} }&
{  \includegraphics[width=0.45\textwidth]{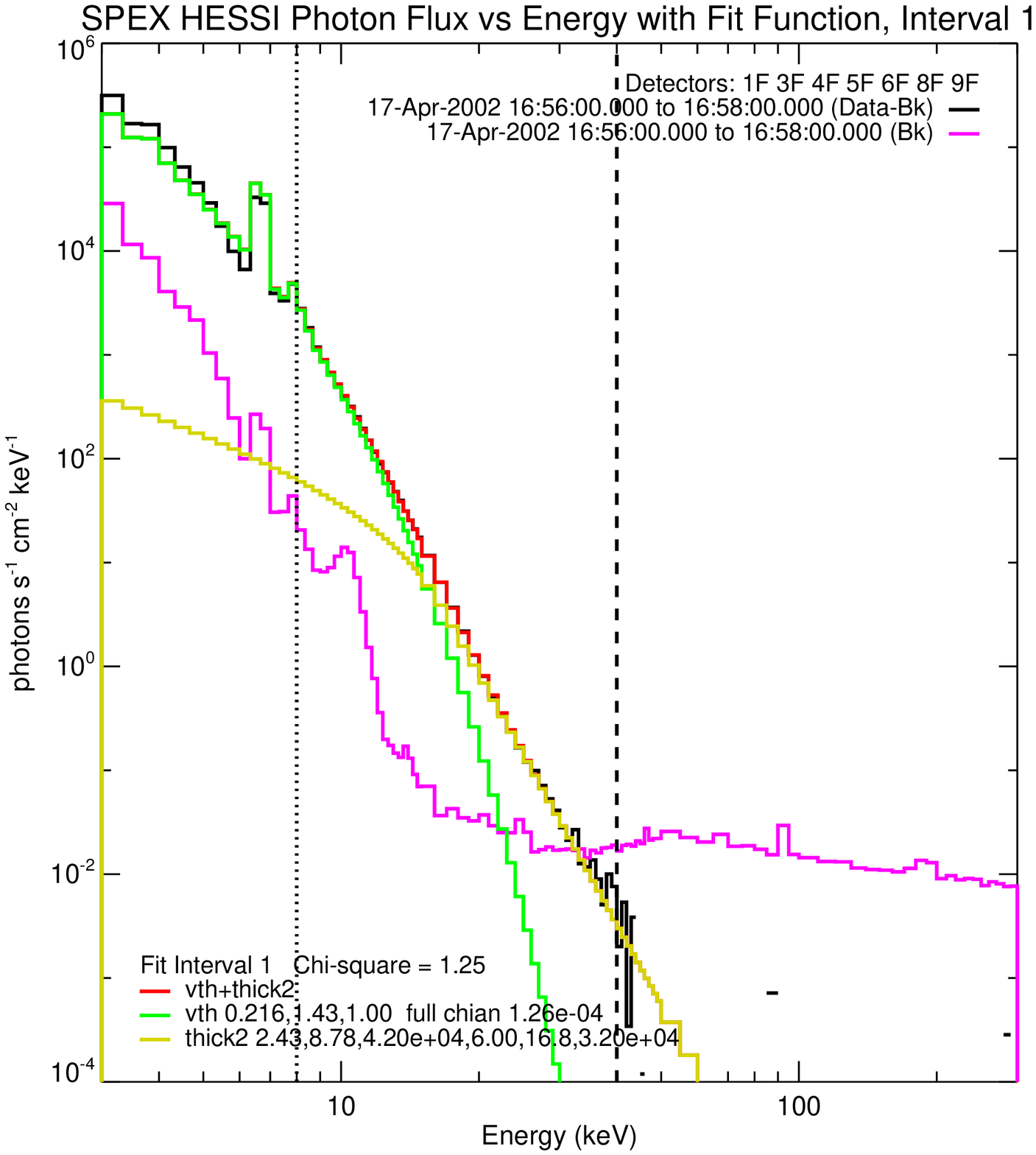}  }\\
\end{tabular}
\caption{ {\it Left panel:} Light curves, in the energy intervals labeled at the top right of the plot, for the flare on 2002~April~17.  The vertical lines delineate the time intervals for Events~6 and~7.  {\it Right panel:} Spectral fit to Event \#7 (16:56:00 - 16:58:00~UT).  The green histogram shows the thermal component of the spectrum (EM = $0.216 \times 10^{49}$~cm$^{-3}$; $T = 1.43$~keV) and the yellow histogram shows the non-thermal thick-target component (transition energy $E_t = 14.8$~keV; spectral index $\delta = 8.78)$.  The red histogram represents the sum of the thermal and nonthermal components and the lilac histogram represents the background.}

\label{light_curve_spectrum}
\end{figure}
\end{center}

The list of events studied is shown in Table~\ref{table:spec}.  In this context, an ``event'' is a time interval during a flare for which spatial and spectral observations are sufficiently good to permit both a determination of the source spatial structure at a variety of energies and the overall spectrum of the hard X-ray emission.  Some flares provide multiple ``events''; other flares only one (see Table~\ref{table:spec}).  For each event, we fit the spatially-integrated hard X-ray emission with an isothermal-plus-power-law form, yielding values (Table~\ref{table:spec}) of the emission measure EM (cm$^{-3}$) and temperature $T$ (keV) of the thermal source, the intensity and spectral index $\delta=\gamma+1$ of the injected nonthermal electron spectrum (corresponding to the hard X-ray spectral index $\gamma$), and $E_t$ (keV), the transition energy between the thermal and nonthermal components.  Straightforward thick-target modeling \citep{brown71} then provides $d{\cal N}/dt$ (s$^{-1}$), the acceleration rate of electrons above (somewhat arbitrary) reference energy $E_0 = 20$~keV.

Parenthetically, we note that all the injected electron spectra are rather steep (the lowest value of $\delta$ is 5.70 [Event \#6], and the typical value is in the range 7 -- 9).  While this may indicate a property of the electron acceleration process in a relatively dense (see below) medium, it may also simply be an observational selection effect -- the relative paucity of high-energy electrons in such steep spectra is consistent with the absence of footpoint emission that such high-energy electrons would produce.

Figure~\ref{light_curve_spectrum} shows the {\em RHESSI} count rate profiles for Events~6~and~7 (16:54:00 - 16:56:00~UT, and 16:56:00 - 16:58:00~UT, respectively, on 2002~April~17) in five different energy channels. We have identified with vertical lines the time intervals for each event.  The right panel shows the spectrum for Event \#7, with the values of the spectral fit parameters provided in the caption (and in Table~\ref{table:spec}).

\begin{center}
\begin{figure}[pht]
\begin{tabular}{cc}
 \includegraphics[width=0.95\textwidth]{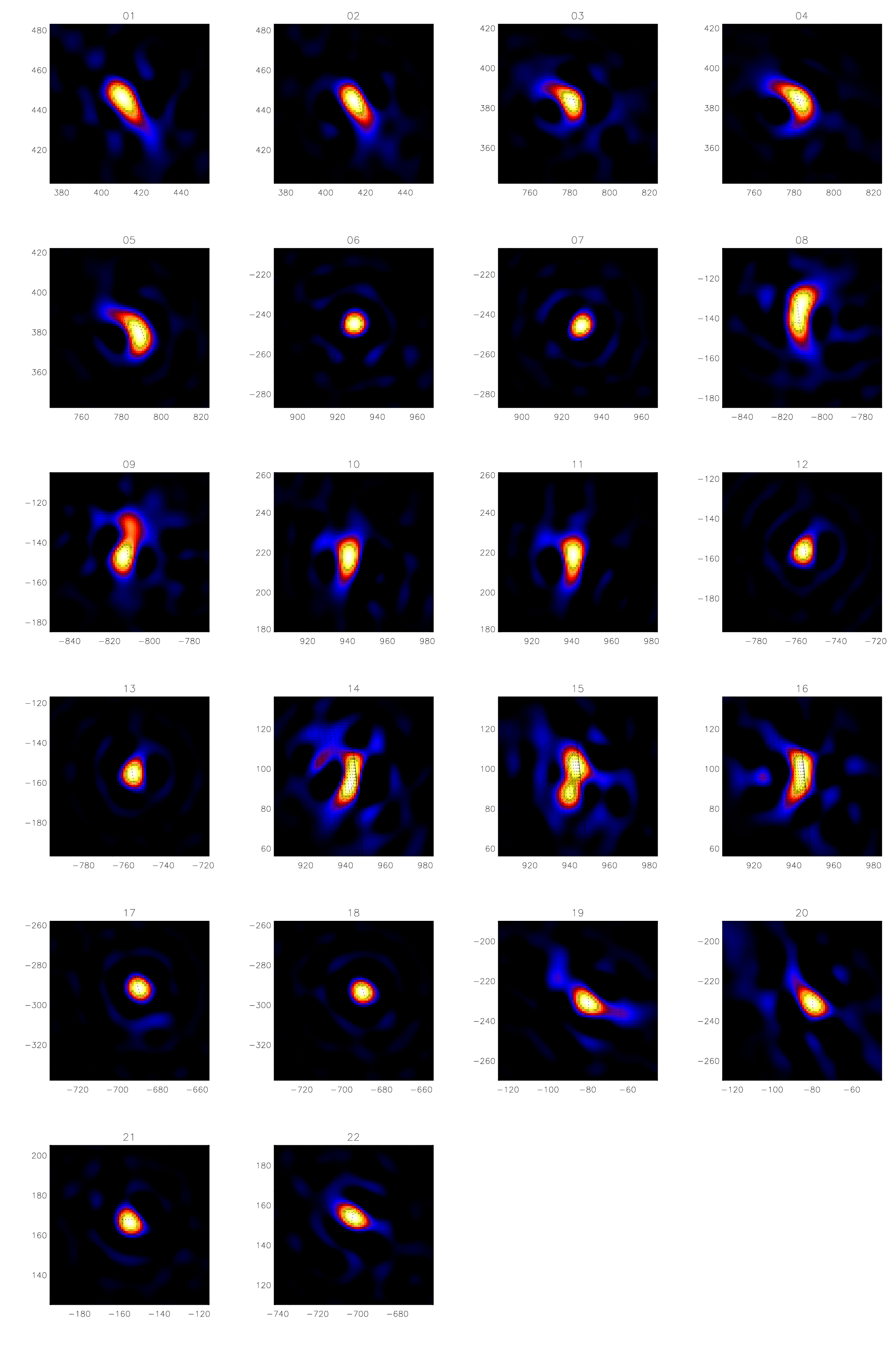}
\end{tabular}
\caption{Mean electron flux maps for each event, in the representative 18-20~keV energy bin.  These maps were obtained by applying the uv-smooth procedure \citep{massone2009} to the electron visibilities \citep{pianaetal07} inferred from the {\em RHESSI} count visibility data.}
\label{fig_maps}
\end{figure}
\end{center}

Electron flux images of each event, in the representative 18-20~keV energy channel, are shown in Figure~\ref{fig_maps}.  These images were produced by performing a spectral inversion on the count visibility data to obtain the corresponding electron visibilities  and then using the uv-smooth algorithm \citep{massone2009} to produce maps of the electron flux \citep[weighted by the line-of-sight column density;][]{pianaetal07}.  Two aspects of this technique are worthy of note.  First, the technique exploits the fact that for the bremsstrahlung process counts of energy $q$ are produced by all electrons with energy $E \ge q$, so that the method yields images at electron energies $E$ up to and beyond the maximum count energy $q_{\rm max}$ observed.  Second, the regularized spectral inversion procedure that produces the electron visibilities results in images that, by construction, vary more smoothly with electron energy $E$ than the ``parent'' count images vary with $q$.  This smooth variation with $E$ greatly facilitates the analysis of the following section.

\section{Analysis}\label{analysis}

Using the electron flux maps, we first calculate the principal longitudinal and lateral directions $s$ and $t$ for the source, using the technique described in \citet{guoetal2012}.  The longitudinal extent of the source external to the acceleration region can be found by considering the standard deviation

\begin{equation}\label{theory}
\sigma(E) = \sqrt{\int_0^\infty s^2 \, F(E,s) \, ds \over \int_0^\infty F(E,s) \, ds} \,\,\, ,
\end{equation}
where $F(E,s)$ is the electron flux spectrum at longitudinal position $s$.

Various physical processes can in principle contribute to the behavior of $F(E,s)$.  Obviously, Coulomb collisions with ambient electrons \citep[e.g.,][]{1978ApJ...224..241E} must be considered.  In addition, various authors \citep[e.g.,][]{1977ApJ...218..306K,1981ApJ...249..817E,2006ApJ...651..553Z} have stressed the need to consider also the Ohmic energy losses associated with driving the beam-neutralizing return current in a resistive medium.  Such energy losses are proportional to the decelerating voltage difference, which is in turn proportional to the beam current.  Return current losses are therefore likely to be significant only in large events \citep[e.g.,][]{1981ApJ...249..817E} with electron acceleration rates $d{\cal N}/dt$ substantially greater than those considered here (Table~\ref{table:spec}).  For similar reasons, we believe that collective plasma effects \citep[see, e.g.,][]{1977ApJ...218..866H,1984ApJ...279..882E} are also likely to be unimportant.  We therefore consider only Coulomb energy losses in the computation of $F(E,s)$.

We therefore consider a cold-target collisional injection model and a target of uniform density $n$ (cm$^{-3}$).  We also neglect pitch angle scattering and dispersion around mean values, which typically affect $F(E,s)$ by factors of order unity \citep{1972SoPh...26..441B,1981ApJ...251..781L}, and we shall address these factors briefly in Section~\ref{results}.  For such a scenario, the form of $F(E,s)$ can be deduced from the one-dimensional continuity and energy loss equations

\begin{equation}\label{assumptions}
F(E) \, dE = F_0(E_0) \, dE_0 \,\,\, ; \qquad {dE \over ds} = - {Kn \over E} \,\,\, .
\end{equation}
(Here $K=2 \pi e^4 \Lambda$, $e$ being the electronic charge and $\Lambda$ being the Coulomb logarithm.)  The solution for $E(s)$ is $E^2 = E_0^2 - 2Kns$, so that $dE_0/dE = E/E_0$ and hence, for a power-law injection spectrum $F_0(E_0) \sim E_0^{-\delta}$,

\begin{equation}\label{fes}
F(E,s) \sim {E \over (E^2 + 2 K n s)^{(\delta + 1)/2}} \,\,\, .
\end{equation}
Substituting the expression~(\ref{fes}) into Equation~(\ref{theory}), we obtain, after some algebra,

\begin{equation}\label{model-electron1}
\sigma(E) = \sqrt{\frac{2}{(\delta - 3)(\delta-5)}} \, {E^2 \over Kn} \,\,\, .
\end{equation}
For a model\footnote{\citet{xuetal08} also discuss a more physically self-consistent model which incorporates the finite density in the acceleration region in the form for $L(E)$.  The corresponding expression for $L(E)$ cannot be expressed as a simple closed form and hence is more complicated to use in a best-fit analysis.  However, this more correct form nevertheless yields results for $L_0$ and $n$ that are comparable to those obtained from the ``tenuous acceleration region'' model used here.} in which electrons are accelerated within a region extending from [$-L_0/2$,$L_0/2$] and injected into an external region with uniform density $n$, we therefore arrive at the relationship between the observed longitudinal source extent $L$ and electron energy $E$:

\begin{equation}\label{model-electron}
{L(E) \over 2} = {L_0 \over 2} + \frac{1}{Kn} \, \sqrt{\frac{2}{(\delta - 3)(\delta-5)}} \, E^2.
\end{equation}
For each event, the values of the acceleration region length $L_0$ and the loop density $n$ are obtained by best-fitting Equation~(\ref{model-electron}) to the inferred form of the variation of loop length $L(E)$ with electron energy $E$ in the predominantly nonthermal domain $E \gapprox E_t$ \citep[see][]{guoetal2012}. The resulting best-fit parameters are presented in Table~\ref{table:params}. From the electron flux images, we also straightforwardly obtain the width (lateral extent) $W$ of the emitting region, which typically exhibits a much smaller variation with energy $E$ than does $L$ \citep[see][]{kohabi11} and can hence be taken as a constant.  From the inferred values of $L_0$, $W$, and $n$, we obtain the volume of the acceleration region

\begin{equation}\label{volume-acc}
V_0= {\pi W^2 L_0 \over 4}
\end{equation}
and the number of particles it contains

\begin{equation}\label{number_of_particles}
{\cal N} = n \, V_0 \,\,\, .
\end{equation}
Values of $V_0$ and ${\cal N}$ are provided for each event in Table~\ref{table:params}.

\subsection{Specific Acceleration Rate}

The {\it specific acceleration rate} (electrons~s$^{-1}$ per electron) is defined \citep{emslie2008AIP} as the ratio of two quantities: $d{\cal N}/dt (\ge E_0)$, the rate of acceleration of electrons beyond energy $E_0$, and ${\cal N}$, the number of particles available for acceleration:

\begin{equation}\label{acceleration-rate}
\eta(E_0) = \frac{1}{\cal{N}} \, \frac{d \cal{N}}{dt} (\ge E_0) \,\,\, .
\end{equation}
The quantity $d{\cal N}/dt (\ge E_0)$ is readily determined by spectral fitting of the spatially-integrated hard X-ray emission -- see Table~\ref{table:spec}.  The quantity ${\cal N}$ can be found from Equation~(\ref{number_of_particles}) -- see values in Table~\ref{table:params}.  We can thus deduce the value of the specific acceleration rate $\eta (E_0)$ in each event; values are given in Table~\ref{table:params}.

\subsection{Filling Factor}

The soft X-ray emission measure EM is related to the plasma density $n$ and the {\it emitting} volume $V_{\rm emit}$ through EM $= n^2 \, V_{\rm emit}$. Given that an emitting region may be composed of a number of discrete emitting subregion (e.g., ``strands,'' ``kernels''), the emitting volume may be equal to or smaller than the total flare volume $V$ estimated from observations of the spatial extent of the source.  The ratio of the emitting volume to the volume $V$ of the observed region that encompasses the emitting region(s) is termed the \emph{filling factor}

\begin{equation}\label{filling-factor}
f = {V_{\rm emit} \over V} = \frac{{\rm EM}}{n^2 V} \,\,\, .
\end{equation}
For each event, we estimated $f$ by using the value of EM from the spectral fit to the thermal portion of the hard X-ray spectrum (Table~\ref{table:spec}), the density $n$ from the fit to Equation~(\ref{model-electron}) (Table~\ref{table:params}), and the observed volume $V$ of the emitting portion of the loop, measured at the transition energy $E_t$ (Table~\ref{table:spec}), the maximum energy at which thermal emission is predominant: $V=(\pi/4) \, W^2 L(E_t)$. Values of $f$ for each event are given in Table~\ref{table:params}.

\section{Results and Conclusions}\label{results}

\begin{deluxetable} {cccccccc}
    \tablewidth{0pt}
    \tabletypesize{\scriptsize}
    \tablecaption{Acceleration Region Characteristics\label{table:params}}
    \tablehead{
    \colhead{Event No.} & \colhead{$L_0$ (arcsec)} & \colhead{$W$ (arcsec)} & \colhead{$V_0$ (100~arcsec$^3$) } & \colhead{$n$ ($10^{11}$~cm$^{-3}$)} & \colhead{${\cal N}$ ($10^{37}$) } & \colhead{$\eta$~(20 keV) ($10^{-3}$~s$^{-1}$) } & \colhead{$f$} }
    \startdata
   1 & $18.6$ & $7.0$  & $7.2$ & $1.5$ & $4.1$ & $6.5$   & $0.45$\\
   2  & $16.3$ & $6.9$  & $6.2$ & $1.4$ & $3.2$ & $14.5$  & $0.83$ \\
   \hline
   3 & $16.7$ & $7.3$ & $7.0$ & $4.4$ & $11.7$ & $4.0$  & $0.04$\\
   4 & $16.6$ & $7.3$ & $7.0$ & $4.8$ & $12.8$ & $7.3$  & $0.11$ \\
   5 & $16.6$ & $8.2$ & $8.7$ & $10.5$ & $34.9$& $3.3$  & $0.03$\\
   \hline
   6 & $11.9$ & $5.9$ & $3.3$ & $4.9$ & $6.0$ & $0.6$   & $0.02$\\
   7 & $10.4$ & $6.0$ & $3.0$ & $1.8$ & $2.0$ & $12.1$  & $0.44$\\
   \hline
   8 & $17.8$ & $6.9$ & $6.4$ & $2.6$ &  $7.1$ & $24.1$ & $0.90$\\	
   9 & $18.8$ & $6.6$ & $6.5$ & $2.9$ &  $7.7$ & $23.1$ & $1.05$  \\
   \hline
   10 & $15.1$ & $6.0$ & $4.2$ & $2.9$ &  $5.4$ & $13.8$  & $0.72$\\
   11 & $16.0$ & $5.7$ & $4.1$ & $1.9$ &  $3.1$ & $27.8$  & $1.95$\\
   \hline
   12 & $10.3$ & $6.6$ & $3.5$ & $5.1$ &  $6.7$ & $4.9$  & $0.08$ \\
   13 & $9.9$ &  $6.5$ & $3.3$ & $4.6$ &  $5.7$ & $4.1$  & $0.18$ \\
   \hline
   14 & $21.5$ & $5.3$ & $4.8$ & $1.5$ & $2.8$ & $1.4$  & $0.13$ \\
   15 & $17.4$ & $6.3$ & $5.4$ & $0.8$ & $1.7$ & $1.7$  & $1.03$ \\
   16 & $17.8$ & $6.4$ & $5.8$ & $2.3$ & $5.1$ & $0.3$  & $0.18$ \\
   \hline
   17 & $11.0$ & $6.2$& $3.3$ &$3.9$ &  $5.0$ & $2.9$  & $0.05$ \\
   18 & $9.9$ & $6.3$& $3.1$ &$3.2$ &  $3.8$ & $7.0$  & $0.22$ \\
   \hline
   19 & $19.9$ & $6.2$ & $6.1$ &$11.1$ & $25.7$ & $13.6$ & $0.02$\\
   20 & $14.5$ & $6.1$ & $4.2$ &$5.2$  & $8.3$ & $23.4$ & $0.10$ \\
   \hline
   21 & $9.9$  & $6.1$ & $2.9$ &$2.2$  & $2.4$ & $16.5$  & $0.53$ \\
   \hline
   22 & $12.4$ & $6.0$ & $3.6$ &$1.7$  & $2.3$ & $5.2$  & $0.26$ \\
   \hline
   Geometric Mean  & 14.5 & 6.4 & 4.7 & 2.9 & 5.4 & 6.0 & 0.20 \\
   $\times/\div$   & 1.3 & 1.1  & 1.4 & 1.9 & 2.2 & 3.4 & 3.9 \\
\enddata
\end{deluxetable}

The values of $L_0$, $W$, $V_0$, $n$, ${\cal N}$, $\eta(20$~keV) and $f$ for each event are presented in Table~\ref{table:params}.  While statistical uncertainties in these values could readily be calculated through a Monte Carlo method in which noise is added to the {\em RHESSI} count visibility data and the process repeated \citep[see][]{guoetal2012}, we have intentionally refrained from doing so here, since the approximations and assumptions used in the model doubtless entail even larger uncertainties.  Instead, we let the scatter of the inferred values of the parameters across the 22 events determine the extent over which the parameters range.  We have calculated (Table~\ref{table:params}) the value of the (geometric) mean value of each quantity and the (multiplicative) uncertainty in this value.  In particular, we obtain $n = (2.9 \times \!\!/\!\div 1.9) \times 10^{11}$~cm$^{-3}$, $f = 0.20 \, \times\!/\!\div 3.9$, and $\eta(20$~keV)~$ = (6.0 \times\!/\!\div 3.4) \times 10^{-3}$~electrons~s$^{-1}$~per ambient electron.
 
Returning to the simplifying assumptions used in determining the form of the electron flux $F(E,s)$ (Equations~[\ref{assumptions}] and~[\ref{fes}]), we note that inclusion of electron trajectories that have a non-zero pitch angle to the guiding magnetic field and/or a guiding magnetic field that is inclined to the longitudinal axis (the direction defining the coordinate $s$) will add a factor $\mu = \overline {\cos \theta}$, where $\theta$ is the angle between the electron velocity vector and the longitudinal direction, to the energy-dependent term in Equation~(\ref{model-electron}). This will result in a decrease (by a multiplicative factor $\mu$) in the inferred density $n$, which in turn, by Equations~(\ref{number_of_particles}), (\ref{acceleration-rate}) and~(\ref{filling-factor}), will increase the values of $f$ and $\eta$ by factors of $1/\mu^2$ and $1/\mu$, respectively.  Inclusion of return current Ohmic energy losses and/or energy losses to waves through collective plasma effects will also decrease the electron penetration depth, leading to further decreases in the inferred value of $n$ and so increases in $f$ and $\eta$.  The values of $f$ and $\eta$ cited above are therefore in all likelihood lower limits.

The inferred values of $f$ are generally somewhat less than unity, with the exception of three events (\#\# 9, 11 and 15), for which $f =$~1.05, 1.95 and 1.03, respectively. Given the uncertainties in the data, the approximations in the analysis method, and the factor of four spread in the inferred values of $f$, neither of these values exceeds unity by an alarming margin.  The mean value of the filling factor obtained is consistent, within a logarithmic standard deviation or so, with unity.  This result, while not entirely surprising, is nevertheless still significant.  It validates the assumption used by many authors \citep[e.g.,][]{emslieetal04} that most of the observed flare volume contains bremsstrahlung-emitting electrons; the degree to which the emission is fragmented (e.g., striated into ``kernels'' or ``strands'' of emission situated within a relatively inert background medium) is quite small.

The inferred mean value of $\eta$(20~keV) $\simeq 5 \times 10^{-3}$~electrons~s$^{-1}$~per ambient electron is broadly consistent with the values reported for a series of extended-loop-source events by \citet{emslie2008AIP}.  It should also be noted that the value of the specific acceleration rate for Event \#4 (the ``midnight flare'' of 2002~April~15) has been determined independently by \citet{torre2012}, who used a continuity equation analysis of the variation of the electron flux spectrum throughout the source. The specific acceleration rate $\eta(20$~keV)$=11 \times 10^{-3}$~s$^{-1}$ obtained by \citet{torre2012} is consistent with the value of $7.3 \times 10^{-3}$~s$^{-1}$ deduced here.

The observationally-deduced value $\eta \simeq 10^{-2}$~s$^{-1}$ implies that all available electrons would be energized and ejected towards the footpoints within a few hundred seconds. This result has significant implications for supply of electrons to the acceleration region, current closure, and the global electrodynamic environment in which electron acceleration and propagation occur \citep[see, e.g.,][]{emslie1995}.

The values of the filling factor $f$ deduced herein are broadly consistent with stochastic acceleration models \citep[e.g.,][Bian et al 2012, ApJ, in press]{petrosian_liu2004, bian2011} which generally involve a near-homogeneous distribution of scattering centers. In addition, the deduced values of the specific acceleration rates $\eta$ are also broadly consistent with such models.  For example, in their study of electron (and proton) acceleration in a turbulent magnetohydrodynamic wave cascade, \citet[][their Figures 6, 7, 9, 10 and 12]{1996ApJ...461..445M} derive values of the volumetric electron acceleration rate $\sim (1.5 - 4) \times 10^8$~cm$^{-3}$~s$^{-1}$ above 20~keV, with the exact value dependent on the assumptions of the various models considered.  In the \citet{1996ApJ...461..445M} model, the background number density is $n=10^{10}$~cm$^{-3}$, so that $\eta \sim (1.5 - 4) \times 10^{-2}$~s$^{-1}$ above 20~keV.

On the other hand, values of the filling factor $f$ close to unity pose significant challenges for particle acceleration models that involve highly localized geometries, such as super-Dreicer acceleration in thin current sheets \citep[see, e.g.,][]{litvinenko2000,turkmani2006pas}.  Turning to the specific acceleration rate in such an acceleration scenario, \citet{heerikhuisen2002paa} find (their equation~[3.17]) a rate of {\it proton} acceleration $d{\cal N}_p/dt \simeq 2 \times 10^{37} \, \sqrt{\zeta}$~s$^{-1}$, where $\zeta$ \citep[$\eta$ in the notation of][]{heerikhuisen2002paa} is the Lundquist number, the ratio of the diffusive to advective terms in the magnetic diffusion equation.  Following \citet{heerikhuisen2002paa}, we take $\zeta = 10^{-8}$, giving a proton acceleration rate $d{\cal N}_p/dt \sim 2 \times 10^{33}$~s$^{-1}$, and we take the number of protons available for acceleration as ${\cal N}_p \sim \varepsilon n \, L^3$, where $\varepsilon \simeq 0.4 \, \sqrt{\zeta} \simeq 4 \times 10^{-5}$ is the angle at the magnetic X-type neutral point at the origin of the acceleration region. With an ambient density $n \simeq 10^{11}$~cm$^{-3}$ and a longitudinal acceleration region extent $L \sim 10^9$~cm, ${\cal N}_p \sim 4 \times 10^{33}$ and so $\eta \simeq 2$~s$^{-1}$.  Although this value is much greater than the values of $\eta$ deduced here (it corresponds to the acceleration of all ambient particles in less than a second), it must be again stressed that the \cite{heerikhuisen2002paa} model refers to highly efficient acceleration of a relatively small number of {\it protons} in a very localized geometry.  We encourage calculations of specific acceleration rates for electrons in such a model, and indeed in all theoretical particle acceleration scenarios.

\acknowledgements JG, AMM and MP have been supported by the EU FP7 Collaborative grant HESPE, grant No. 263086; AGE was supported by NASA Grant NNX10AT78J. The authors thank the referee for helpful comments and Richard Schwartz, Gabriele Torre, Eduard Kontar and Federico Benvenuto for useful discussions.

\bibliographystyle{apj}
\bibliography{guo_age}

\begin{thebibliography}{27}
\expandafter\ifx\csname natexlab\endcsname\relax\def\natexlab#1{#1}\fi

\bibitem[{{Bian} {et~al.}(2011){Bian}, {Kontar}, \& {MacKinnon}}]{bian2011}
{Bian}, N.~H., {Kontar}, E.~P., \& {MacKinnon}, A.~L. 2011, A\& A, 535, A18

\bibitem[{{Brown}(1971)}]{brown71}
{Brown}, J.~C. 1971, Sol.~Phys., 18, 489

\bibitem[{{Brown}(1972)}]{1972SoPh...26..441B}
---. 1972, \solphys, 26, 441

\bibitem[{{Emslie}(1978)}]{1978ApJ...224..241E}
{Emslie}, A.~G. 1978, \apj, 224, 241

\bibitem[{{Emslie}(1981)}]{1981ApJ...249..817E}
---. 1981, \apj, 249, 817

\bibitem[{{Emslie} \& {H\'enoux}(1995)}]{emslie1995}
{Emslie}, A.~G., \& {H\'enoux}, J.-C. 1995, \apj, 446, 371

\bibitem[{{Emslie} {et~al.}(2008){Emslie}, {Hurford}, {Kontar}, {Massone},
  {Piana}, {Prato}, \& {Xu}}]{emslie2008AIP}
{Emslie}, A.~G., {Hurford}, G.~J., {Kontar}, E.~P., {Massone}, A.~M., {Piana},
  M., {Prato}, M., \& {Xu}, Y. 2008, in American Institute of Physics
  Conference Series, Vol. 1039, American Institute of Physics Conference
  Series, ed. {G.~Li, Q.~Hu, O.~Verkhoglyadova, G.~P.~Zank, R.~P.~Lin, \&
  J.~Luhmann }, 3--10

\bibitem[{{Emslie} {et~al.}(2004){Emslie}, {Kucharek}, {Dennis}, {Gopalswamy},
  {Holman}, {Share}, {Vourlidas}, {Forbes}, {Gallagher}, {Mason}, {Metcalf},
  {Mewaldt}, {Murphy}, {Schwartz}, \& {Zurbuchen}}]{emslieetal04}
{Emslie}, A.~G., {Kucharek}, H., {Dennis}, B.~R., {Gopalswamy}, N., {Holman},
  G.~D., {Share}, G.~H., {Vourlidas}, A., {Forbes}, T.~G., {Gallagher}, P.~T.,
  {Mason}, G.~M., {Metcalf}, T.~R., {Mewaldt}, R.~A., {Murphy}, R.~J.,
  {Schwartz}, R.~A., \& {Zurbuchen}, T.~H. 2004, J.~Geophys.~Res. (Space
  Physics), 109, 10104

\bibitem[{{Emslie} \& {Smith}(1984)}]{1984ApJ...279..882E}
{Emslie}, A.~G., \& {Smith}, D.~F. 1984, \apj, 279, 882

\bibitem[{{Guo} {et~al.}(2012){Guo}, {Emslie}, {Kontar}, {Benvenuto},
  {Massone}, \& {Piana}}]{guoetal2012}
{Guo}, J., {Emslie}, A.~G., {Kontar}, E.~P., {Benvenuto}, F., {Massone}, A.~M.,
  \& {Piana}, M. 2012, A\&A, in press

\bibitem[{Heerikhuisen {et~al.}(2002)Heerikhuisen, Litvinenko, \&
  Craig}]{heerikhuisen2002paa}
Heerikhuisen, J., Litvinenko, Y., \& Craig, I. 2002, ApJ, 566, 512

\bibitem[{{Hoyng} \& {Melrose}(1977)}]{1977ApJ...218..866H}
{Hoyng}, P., \& {Melrose}, D.~B. 1977, \apj, 218, 866

\bibitem[{{Knight} \& {Sturrock}(1977)}]{1977ApJ...218..306K}
{Knight}, J.~W., \& {Sturrock}, P.~A. 1977, \apj, 218, 306

\bibitem[{{Kontar} {et~al.}(2011){Kontar}, {Hannah}, \& {Bian}}]{kohabi11}
{Kontar}, E.~P., {Hannah}, I.~G., \& {Bian}, N.~H. 2011, ApJL, 730, L22

\bibitem[{Krucker {et~al.}(2008)Krucker, Battaglia, Cargill, Fletcher, Hudson,
  MacKinnon, Masuda, Sui, Tomczak, Veronig, {et~al.}}]{krucker08}
Krucker, S., Battaglia, M., Cargill, P., Fletcher, L., Hudson, H., MacKinnon,
  A., Masuda, S., Sui, L., Tomczak, M., Veronig, A., {et~al.} 2008, A\& A
  Review, 16, 155

\bibitem[{{Leach} \& {Petrosian}(1981)}]{1981ApJ...251..781L}
{Leach}, J., \& {Petrosian}, V. 1981, \apj, 251, 781

\bibitem[{{Litvinenko} \& {Craig}(2000)}]{litvinenko2000}
{Litvinenko}, Y.~E., \& {Craig}, I.~J.~D. 2000, ApJ, 544, 1101

\bibitem[{{Massone} {et~al.}(2009){Massone}, {Emslie}, {Hurford}, {Prato},
  {Kontar}, \& {Piana}}]{massone2009}
{Massone}, A.~M., {Emslie}, A.~G., {Hurford}, G.~J., {Prato}, M., {Kontar},
  E.~P., \& {Piana}, M. 2009, ApJ, 703, 2004

\bibitem[{{Miller} {et~al.}(1996){Miller}, {Larosa}, \&
  {Moore}}]{1996ApJ...461..445M}
{Miller}, J.~A., {Larosa}, T.~N., \& {Moore}, R.~L. 1996, \apj, 461, 445

\bibitem[{{Petrosian} \& {Liu}(2004)}]{petrosian_liu2004}
{Petrosian}, V., \& {Liu}, S. 2004, ApJ, 610, 550

\bibitem[{{Piana} {et~al.}(2007){Piana}, {Massone}, {Hurford}, {Prato},
  {Emslie}, {Kontar}, \& {Schwartz}}]{pianaetal07}
{Piana}, M., {Massone}, A.~M., {Hurford}, G.~J., {Prato}, M., {Emslie}, A.~G.,
  {Kontar}, E.~P., \& {Schwartz}, R.~A. 2007, ApJ, 665, 846

\bibitem[{{Sui} {et~al.}(2004){Sui}, {Holman}, \& {Dennis}}]{suetal04}
{Sui}, L., {Holman}, G.~D., \& {Dennis}, B.~R. 2004, ApJ, 612, 546

\bibitem[{{Torre} {et~al.}(2012){Torre}, {Pinamonti}, {Emslie}, {Guo},
  {Massone}, \& {Piana}}]{torre2012}
{Torre}, G., {Pinamonti}, N., {Emslie}, A.~G., {Guo}, J., {Massone}, A.~M., \&
  {Piana}, M. 2012, ApJ, 751, 129

\bibitem[{Turkmani {et~al.}(2006)Turkmani, Cargill, Galsgaard, Vlahos, \&
  Isliker}]{turkmani2006pas}
Turkmani, R., Cargill, P., Galsgaard, K., Vlahos, L., \& Isliker, H. 2006,
  A\&A, 449, 749

\bibitem[{Veronig \& Brown(2004)}]{vebr04}
Veronig, A., \& Brown, J. 2004, ApJL, 603, L117

\bibitem[{{Xu} {et~al.}(2008){Xu}, {Emslie}, \& {Hurford}}]{xuetal08}
{Xu}, Y., {Emslie}, A.~G., \& {Hurford}, G.~J. 2008, ApJ, 673, 576

\bibitem[{{Zharkova} \& {Gordovskyy}(2006)}]{2006ApJ...651..553Z}
{Zharkova}, V.~V., \& {Gordovskyy}, M. 2006, \apj, 651, 553

\end{thebibliography}

\end{document}